\documentclass[pra,jor,amsmath,amssymb,reprint]{revtex4-2}

\usepackage{graphicx}
\usepackage{dcolumn}
\usepackage{bm}
\usepackage{braket}

\begin{document}

\preprint{AIP/123-QED}

\title{Ultrafast Two-electron Orbital Swap in Li Initiated by Attosecond pulses}
\author{Hui Jiang$^{1}$}
\author{Zhao-Han Zhang$^{1}$}
\author{Yang Li$^{1}$}
\author{Camilo Ruiz$^{2}$}
\author{Feng He$^{1,3}$}\email{fhe@sjtu.edu.cn}
	
\affiliation{
	$^1$Key Laboratory for Laser Plasmas (Ministry of Education) and School of Physics and Astronomy,
	Collaborative innovation center for IFSA (CICIFSA), Shanghai Jiao Tong University, Shanghai 200240, China\\
	$^2$Instituto Universitario de F\'{i}sica Fundamental y Matem\'aticas, Universidad de Salamanca, Plaza de la Merced s/n, 37008 Salamanca, Spain\\
	$^3$CAS Center for Excellence in Ultra-intense Laser Science, Shanghai 201800, China
}

\date{\today}
	
\begin{abstract}
 A universal mechanism of ultrafast two-electron orbital swap is discovered through
 two-photon sequential double ionization of Li. After a $1s$ electron in Li is ionized by absorbing
 an EUV photon, the other two bound electrons located on two different shells have either parallel or
 antiparallel spin orientations. In the latter case, these two electrons are in the
 superposition of the singlet and triplet states with different energies, forming a quantum beat
 and giving rise to the two-electron orbital swap with a period of several hundred attoseconds.
 The orbital swap mechanism can be used to manipulate the spin polarization of photoelectron pairs
 by conceiving the attosecond-pump attosecond-probe strategy, and thus serves as a knob 
 to control spin-resolved multielectron ultrafast dynamics.
\end{abstract}
	
\maketitle

\section{Introduction}

Chemical reactions fundamentally rely on dynamics of charged particles in the time scale of
femtoseconds or even attoseconds \cite{Nisoli17}. 
The advent of attosecond pulses \cite{Chini14} makes it possible to capture
ultrafast electron dynamics with unprecedented time resolutions. By tracing electron movies, one is able to discover
mechanisms governing ultrafast dynamics. Implementation of these mechanisms to steer ultrafast
reactions under strong lasers has been a long aim in attosecond physics \cite{attoscience}.
An important way of steering ultrafast reactions would be to launch an electron to multiple coherent states,
forming quantum beats that initiate ultrafast charge density oscillations, which may be used to probe electron-electron
correlations \cite{Hu06} and control electron localization in molecular dissociation \cite{He07,Sansone10}.

Although a series of ultrafast processes have been explored in multielectron systems,
we are still far from a full understanding of ultrafast dynamics in atoms with more than
two electrons in strong laser fields. The main obstacle is the lack of fully correlated 
three-dimensional 
quantum calculations which are beyond the capabilities of current computers. A circuitous route is
to use the single-active-electron approximation, which is very successful especially when
only a valence electron contributes to the main dynamics \cite{SAE1,SAE2}. However, if
a multi-electron atom is exposed to high-frequency light sources such as X-ray
free-electron lasers \cite{XFEL1,XFEL2} and attosecond pulses \cite{XFELatto}, an inner-shell
electron may be preferentially kicked off, and the shell structure sustained by the Pauli exclusion
principle becomes unstable. Therefore, electron correlations become important in multi-electron ultrafast processes \cite{Auger,hole1,hole2,hole3}.
As the simplest open-shell atom, Li is used as a benchmark to study multielectron 
effects. Numerical models based on the close-coupling method \cite{Rmatrix,CCC,TDCC1,TDCC2,TDCC3} can provide reliable 
photoionization cross sections of Li \cite{exp1,exp2,exp4}.
The directly numerical simulations of one-dimensional time-dependent Schr\"odinger equation (TDSE) 
have been used to study spin-resolved strong field ionization \cite{1D-TDSE1,1D-TDSE2,jakub18}.
Previous works \cite{cross-s1,cross-s2,angular1,Efimov2019} showed that the electron spin configuration plays a major role for photoionization dynamics in multi-electron systems.
However, the main mechanism behind the spin-related conclusions in these works is limited to the photoionization cross
sections in which the scattering dynamics between electrons will be influenced by the spin configurations (the Pauli exclusion
principle).
One may expect more universal and fundamental mechanisms that can be used to perform ultrafast spin-resolved control for multi-electron dynamics.
	
	\begin{figure*}
		\centering
		\includegraphics[width=0.7\textwidth]{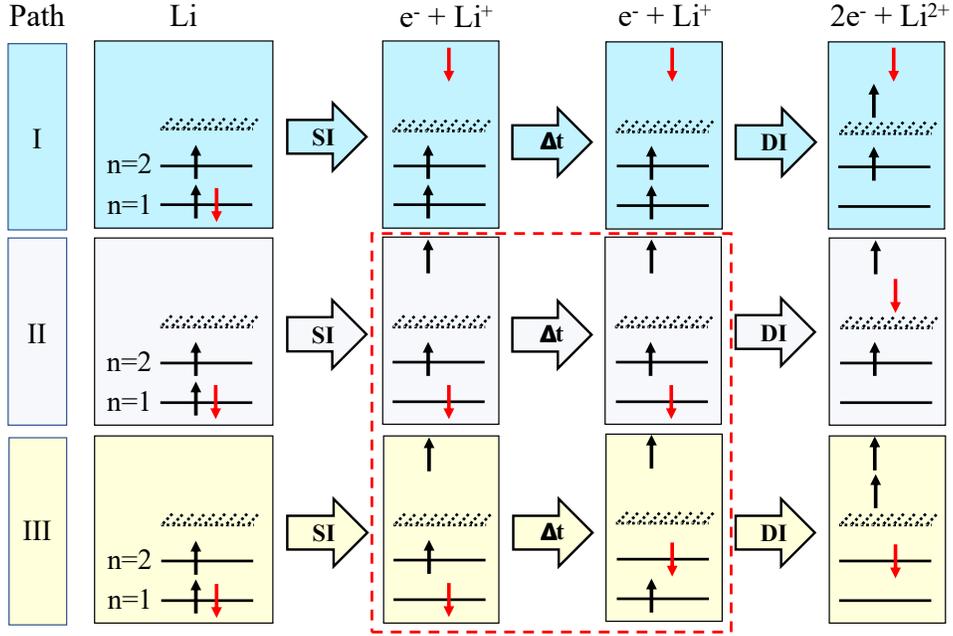}
		\caption{Schematic diagram of three paths for sequential two-photon
			double ionization of two inner-shell electrons. 
			In path-I, an upper-spin inner-shell escapes, leaving the two bound 
			electrons with same spin orientations 
			In path-$\mathrm{\uppercase\expandafter{\romannumeral2}}$
			and path-$\mathrm{\uppercase\expandafter{\romannumeral3}}$, Li$^{+}$ undergoes orbital
			swap (surrounded by the red dotted line).}
		\label{F1}
	\end{figure*}

    In this work, we study the spin-selective ionization of Li in attosecond EUV pulses and discover a fundamental ultrafast process of orbital swap 
    of bound electrons in Li$^+$. We numerically simulate the TDSE including three active electrons initially being prepared 
    in the ground state ($S=1/2$, $M_S=1/2$).
    For atoms in EUV fields, an inner-shell electron always preferentially absorbs an EUV photon if the photon
    energy is big enough. Once an inner-shell electron absorbs an EUV photon, according to its spin orientation, the residual electrons
	in Li$^+$ could be either in the parallel or anti-parallel spin state, as depicted in
	Fig. \ref{F1}. In the former case, the two identical electrons stay in the triplet state ($S=1$, $M_S=1$).
	However, in the latter case of the anti-parallel spin state,
	the two-electron state cannot be fully described by a single
	configuration, and will actually evolve as a superposition of the singlet ($S=0$, $M_S=0$)
	and the triplet ($S=1$, $M_S=0$) states as time flows. Such a quantum beat leads to the periodical
	orbital swap of the two electrons in Li$^+$. In the later time, another inner-shell electron may absorb
	the second photon and gets freed. There are two directly outputs in sequential two-photon double ionization of two inner-shell electrons according to the orbital swap.
	First, the time delay of absorbing two photons depends on the spin orientations of the first photoelectron. Second, the spin orientations of the second 
	photoelectron can be selected, and thus it is possible to produce spin-polarized electron pairs with an attosecond pump-probe strategy.
	In contrast to previous studies about the spin selectivity of the photoelectron \cite{circular1,circular2,circular3,circular4,twocolor}, our demonstration of spin-polarized electron pairs does not rely on the spin-orbit interaction \cite{spinorbit} and paves a new way to perform the spin control of photoelectrons with the orbital swap mechanism. In essence, the mechanism discovered in Li is general and exists in other open-shell atoms as well. 
    Such a mechanism makes it possible to control spin-resolved ultrafast processes induced by strong laser fields.

\section{Theory}
	
	In Li, the two inner-shell electrons couple to the singlet spin state \cite{cross-s2}, which then 
	couples with the third spin-up electron, forming the state ($S=1/2, M_S=1/2$) where we have 
	assumed that the outer-shell electron is spin-up before introducing lasers. The three-electron wave function
	satisfies the exchange asymmetry and is written as
	\begin{equation}
		\begin{aligned}
			\Psi(q_1,q_2,q_3)=\mathcal{A}\Big\{& \frac{1}{\sqrt{2}}\left [\alpha(1) \beta(2) \alpha(3)-\beta(1) \alpha(2) \alpha(3)\right ]\\
			& \psi\left(x_{1}, x_{2}, x_{3}\right)\Big\}.
		\end{aligned}
		\label{eq1}
	\end{equation}
	Here, $q_{i}$ is the spin-spatial coordinate, $x_{i}$ is the spatial coordinate and $\psi$ is the
	spatial wave function, $\mathcal{A}$ the antisymmetrization operator, and $\alpha(i)$ and $\beta(i)$
	represent spin-up and spin-down states, respectively. After some algebra, the wave function
	can be reformulated as
	\begin{equation}
		\Psi(q_1,q_2,q_3)=\Theta\Big\{\alpha(1)\alpha(2)\beta(3)\Phi_{\alpha\alpha\beta}(x_1,x_2,x_3)\Big\},
		\label{eq2}
	\end{equation}
	where $\Theta$ performs the cyclic sum $\Theta f(1,2,3)=[f(1,2,3)+f(2,3,1)+f(3,1,2)]/\sqrt{3}$ for arbitrary $f$.
	$\Phi_{\alpha\alpha\beta}(x_1,x_2,x_3)$ is the spatial wave function associated with the spin base $\alpha(1)\alpha(2)\beta(3)$,
	and it is antisymmetric under the exchange of $x_1$ and $x_2$. $\Phi_{\alpha\alpha\beta}(x_1,x_2,x_3)$
	can be explicitly written as
	\begin{equation}
		\begin{aligned}
			\Phi_{\alpha\alpha\beta}(x_1,x_2,x_3)=\frac{1}{2}\left[\psi\left(x_{2}, x_{3}, x_{1}\right)+\psi\left(x_{3}, x_{2}, x_{1}\right)\right. \\
			\left. -\psi\left(x_{1}, x_{3}, x_{2}\right)-\psi\left(x_{3}, x_{1}, x_{2}\right) \right].
			\label{eq3}
		\end{aligned}
	\end{equation}
	The wave function $\Phi_{\alpha\alpha\beta}$ shows that $e_2^{\uparrow},
	e_3^{\downarrow}$ (or $e_1^{\uparrow}, e_3^{\downarrow}$) are in the inner shell
	and the first (or last) two terms of the spatial wave function are symmetric under the exchange 
	of $x_2$ (or $x_1$) and $x_3$, and $e_1^{\uparrow}$ (or $e_2^{\uparrow}$) is in the outer shell.

The ab initio simulation of the TDSE including three active electrons in full dimensions demands
extraordinarily heavy calculation resources, which is out of the capability of the most advanced
computers in the world. To make it feasible, in this study, we use a linearly polarized EUV
pulse to interact with Li, and confine the electron movement along the laser polarization direction.
The simulation using such a reduced-dimensionality model is still heavy. In one dimension, 
the $n=1$ and $n=2$ orbitals can be represented as $1s$ and $2p$. Therefore, 
we express the main configuration of the ground state based on Eq. (\ref{eq3}) as
	\begin{equation}
		\begin{aligned}
			\sqrt{2}\Phi_{\alpha \alpha \beta}\left(x_{1}, x_{2}, x_{3}\right)=\langle x_{2}, x_{3}|1 s^{2}\,^1 S\rangle \langle x_{1} | 2 p\rangle\\
			-\langle x_{1}, x_{3}|1 s^{2}\,^1 S\rangle \langle x_{2}| 2 p\rangle.
		\end{aligned}
		\label{eq4}
	\end{equation}
	Though the one-dimension model cannot describe the ionization process accurately and the electron-electron
	correlation is overestimated, however, the exchange symmetry characters of the system are preserved. Therefore, 
the reduced one-dimension model can grasp the central physics and qualitatively describe the dynamics we 
discussed in this paper.
	
In Eq. (\ref{eq2}), different spin bases are orthogonal by definition, and no transition between different
	spin bases occurs within the dipole approximation. Without loss of generality, we simulate
	the TDSE under the spin base $\alpha(1)\alpha(2)\beta(3)$ (atomic units are used throughout unless stated otherwise)
	\begin{equation}
		{\rm{i}} \frac{\partial}{\partial t} \Phi_{\alpha\alpha\beta}(x_1,x_2,x_3;t)=[H_0+W(t)]\Phi_{\alpha\alpha\beta}(x_1,x_2,x_3;t),
		\label{eq_TDSE}
	\end{equation}
	where the field-free Hamiltonian is
	\begin{equation}
		\begin{aligned}
			H_0=&\sum_{i=1}^3 \left(-\frac{1}{2}\frac{\partial^2}{\partial x_i^2}-\frac{3}{\sqrt{x_i^2+s^2}}\right) \\ &+\sum_{i=1}^3\sum_{j=1}^{i-1}\frac{1}{\sqrt{(x_i-x_j)^2+s^2}}
			\label{eq_H}
		\end{aligned}
	\end{equation}
	with $s$ the soft-core parameter to adjust the ground state energy in the model. 
	Within the dipole approximation, the laser-Li coupling is expressed as
	\begin{equation}
		W(t)=\sum_{i=1}^3\left[-{\mathrm{i}} A(t)\frac{\partial}{\partial x_i}\right]
	\end{equation}
	with $A(t)$ the laser vector potential.
	The ground state is obtained by imaginary-time propagation of the field free Schr\"odinger equation 
	while constraining the trial wave function to have the same exchange
	property as $\Phi_{\alpha\alpha\beta}$ \cite{imag-time},
	and the real-time propagation is performed by the Crank-Nicolson method \cite{Crank}.
	In each dimension, the box covers the area of $[-440, 440]$ a.u.. This size is
	big enough to hold all the two-photon double ionization events during the whole simulation.
	The spatial steps are $\Delta x_1=\Delta x_2=\Delta x_3=0.2$ a.u., and the time
	step is $\Delta t=0.05$ a.u. Simulation convergence has been tested by using smaller time-spatial grids
	and same results are obtained. The ground state energy $-7.477$ a.u. is ensured by setting $s=0.504$.
	In our calculations, the laser vector potential is written as
	\begin{equation}
		A(t)=A_0\sin^2(\pi t/\tau_{\rm EUV})\sin(\omega t),~t\in[0, \tau_{\rm EUV}].
		\label{laser}
	\end{equation}
	The central frequency is fixed at $\omega=5$ a.u., and the laser has an intensity
	$4.0 \times 10^{16} \mathrm{~W} / \mathrm{cm}^{2}$. Note that the laser intensity is
	not crucial for the mechanism we discover in this study, and we use the high intensity 
	only for obtaining a better signal-to-noise ratio. In potential experiments in the near
	future, the EUV intensity can be chosen according to detect enough two-photon double
	ionization signals. 
	
	\begin{center}
		\begin{table}
			\renewcommand\arraystretch{1.3}
			\caption{\label{table1}
				Energy levels of one-dimensional targets.}
			\begin{ruledtabular}
				\begin{tabular}{ccccc}
					Target  & Configuration & S & Energy (a.u.) \\
					\hline
					Li & $1s^22p$ & $\frac{1}{2}$ & -7.477 \\
					Li$^{+}$ & $1s^2$ & 0 & -7.098 \\
					Li$^{+}$ & $1s2p$ & 1 & -5.472 \\
					Li$^{+}$ & $1s2p$ & 0 & -5.207 \\
					Li$^{2+}$ & $1s$ & $\frac{1}{2}$ & -4.267 \\
					Li$^{2+}$ & $2p$ & $\frac{1}{2}$ & -2.027 \\
				\end{tabular}
			\end{ruledtabular}
		\end{table}
	\end{center}
	
	Due to the electron-electron correlation, the ground state of Li $\Phi_{\alpha\alpha\beta}(x_1,x_2,x_3)$ is not the
	direct product of three single-particle states $\psi_{n_1}(x_1)\psi_{n_2}(x_2)\psi_{n_3}(x_3)$, but can
	be written as $\sum\limits_{n_1,n_2,n_3}C_{n_1,n_2,n_3}\psi_{n_1}(x_1)\psi_{n_2}(x_2)\psi_{n_3}(x_3)$
	where $\psi_{n_j}(x_j)$	is the eigenstate of one-dimensional Li$^{2+}$, and $C_{n_1,n_2,n_3}$ is the amplitude
	of $\psi_{n_1}(x_1)\psi_{n_2}(x_2)\psi_{n_3}(x_3)$ and it shows the main configuration of the ground state in the single-particle bases.
	$\psi_{n_j}(x_j)$ can be obtained by solving the stationary Schr\"odinger equation for Li$^{2+}$
	\begin{equation}
		\left [-\frac 1 2 \frac{\partial^2}{\partial x_j^2}-\frac{3}{\sqrt{x_j^2+s^2}}\right ]\psi_{n_j}(x_j)=E_{n_j}\psi_{n_j}(x_j).
		\label{1d}
	\end{equation}
	The probability for $\Phi_{\alpha\alpha\beta}(x_1,x_2,x_3)$ on $\psi_{n_1}(x_1)\psi_{n_2}(x_2)\psi_{n_3}(x_3)$
	is $|C_{n_1,n_2,n_3}|^2$, as shown in the most right column in Table \ref{table2}.
    Similarly, by removing one electron in Eq. (\ref{eq_TDSE}) and following the same procedure as that done
    for Eq. (\ref{1d}), one may calculate the Eigen energies for Li$^+$. 
	In Table \ref{table1}, we show the energy levels of our model for Li, Li$^+$, and Li$^{2+}$ in different states.
	In the last four rows in the Table, "$\times$" means no constraint on this spatial wave function.
	The data in Table \ref{table2} demonstrate that $e_3^{\downarrow}$ must be on the inner shell
	(occupy the $1s$ orbit), and the other two spin-up electrons are located on
	different shells (mainly on $1s$ and $2p$ states).  Thus, the expression of Eq. (\ref{eq4}) makes sense.

	\begin{center}
		\begin{table}
			\renewcommand\arraystretch{1.3}
			\caption{\label{table2}
				Projection of $\Phi_{\alpha\alpha\beta}(x_1,x_2,x_3)$ onto single-particle states.}
			\begin{ruledtabular}
				\begin{tabular}{ccccc}
					$\ket{\psi_{n_1}}$ &$\ket{\psi_{n_2}}$ & $\ket{\psi_{n_3}}$ & $|C_{n_1,n_2,n_3}|^2$ \\
					\hline
					$\ket{1s}$  & $\ket{1s}$ & $\ket{1s}$ & $<10^{-10}$ \\
					$\ket{1s}$  & $\ket{2p}$ & $\ket{1s}$ & 0.367 \\
					$\ket{2p}$  & $\ket{1s}$ & $\ket{1s}$ & 0.367 \\
					$\ket{2p}$  & $\ket{2p}$ & $\ket{1s}$ & $<10^{-10}$ \\
					$\ket{1s}$  & $\ket{1s}$ & $\ket{2p}$ & $<10^{-10}$ \\
					$\ket{1s}$  & $\ket{2p}$ & $\ket{2p}$ & $<10^{-10}$ \\
					$\ket{2p}$  & $\ket{1s}$ & $\ket{2p}$ & $<10^{-10}$ \\
					$\ket{2p}$  & $\ket{2p}$ & $\ket{2p}$ & $<10^{-10}$ \\
					$\times$ & $\times$ & $\ket{1s}$ & 0.992 \\
					$\times$  & $\times$ & $\ket{2p}$ & 0.004 \\
					$\ket{1s}$  & $\times$ & $\times$ & 0.493 \\
					$\ket{2p}$  & $\times$ & $\times$ & 0.376 \\
				\end{tabular}
			\end{ruledtabular}
		\end{table}
	\end{center}
	
	\section{Single ionization of Li}
	
	\begin{figure*}
		\centering
		\includegraphics[width=0.8\textwidth]{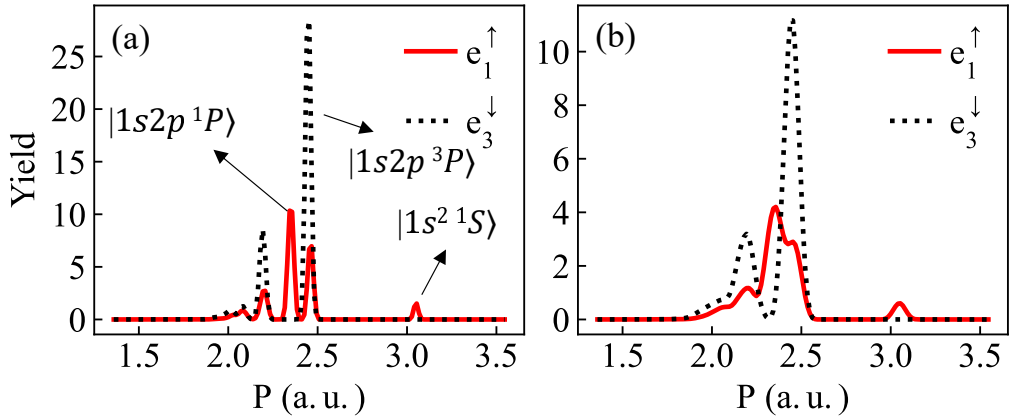}
		\caption{Photoelectron momentum spectra (only the positive parts) for the single ionization of Li
			when the laser pulse duration is (a) 100 a.u. and (b) 38 a.u.. Each panel is normalized by the single ionization rate of $e_1^{\uparrow}$.
		}
		\label{FS1}
	\end{figure*}
	
	The study of single ionization can lay a foundation for our research on double ionization.
	By simulating the TDSE including three active electrons, we may trace the wave function evolution with time.
	At the end of the calculation, we collect the wave function in the area 
	$\left(|x_1|>30\ \text{a.u.}, |x_2|<30\ \text{a.u.}, |x_3|<30\ \text{a.u.} \right)$,
	or $\left(|x_1|<30\ \text{a.u.}, |x_2|>30\ \text{a.u.}, |x_3|<30\ \text{a.u.} \right)$
	or $\left(|x_1|<30\ \text{a.u.}, |x_2|<30\ \text{a.u.}, |x_3|>30\ \text{a.u.} \right)$, and
	transform them to momentum representation to get the momentum distribution.
	Figures \ref{FS1} (a) and (b) show the photoelectron momentum distributions in the positive half spaces
	using the laser pulse durations of 100 a.u. and 38 a.u., respectively. 
	In both panels, the solid and dashed curves are
	for $e_1^{\uparrow}$ and $e_3^{\downarrow}$, respectively.
	The dashed curve in Fig. \ref{FS1}(a) mainly presents two peaks. This is understandable with the help
	of Table \ref{table2}. The emission of $e_3^{\downarrow}$, which must be in the $1s$ shell, leaves
	Li$^+$ in the $\ket{1s2p~^3P}$ state or even higher excited states. If the ion is in the $\ket{1s2p~^3P}$
	state, the momentum of the emitting electron is $\sqrt{2(\omega-I_P^A)}$ a.u., where $I_P^A$ is the
	potential to ionize Li to Li$^+$ in the $\ket{1s2p~^3P}$ state. As shown in Table \ref{table1}, $I_P^A=2.005$ a.u.,
	and thus the momentum is equal to 2.45 a.u., which is coincident with the highest peak of
	the dashed line in Fig. \ref{FS1}(a).
	Alternatively, after $e_3^{\downarrow}$ absorbs a photon, it may excite another electron via electron-electron correlation, i.e.,
	the single photon energy is shared between $e_3^{\downarrow}$ and the outer-shell electron.
	Such a process contributes the momentum peak at 2.2 a.u..
	
	The emission of $e_1^{\uparrow}$ brings more complex structures. $e_1^{\uparrow}$ can be in either the $1s$ or $2p$
	shell, and thus has different ionization potentials. For $e_1^{\uparrow}$ in the $2p$ shell, its
	ionization potential is $I_p^1=0.379$ a.u. in our one-dimensional model,
	and thus the photoionization gives the momentum $\sqrt{2(\omega-I_p^1)}=3.04$ a.u.. Note that this peak is missing in the
	dashed curve since $e_3^{\downarrow}$ cannot be in the outer shell. On the other hand, if $e_1^{\uparrow}$ is in the inner shell,
	the emission of $e_1^{\uparrow}$ leaves Li$^+$ in the superposition of $\ket{1s2p~^3P}$ and $\ket{1s2p~^1P}$.
	Since $\ket{1s2p~^3P}$ and $\ket{1s2p~^1P}$ are spatially exchange symmetry and antisymmetry, the electron-electron
	correlation is different and the two states have different energies. Indeed, the two peaks at 2.34 a.u. and 2.45 a.u. correspond
	to Li$^+$ in the $\ket{1s2p~^1P}$ and $\ket{1s2p~^3P}$ states, respectively.
	As one can clearly see, the peak at 3.04 a.u. is much lower than others, which is
	due to the fact that the inner-shell electron is preferential to absorb highly energetic photons \cite{cross-s1,cross-s2}.
	When a pulse with a short duration is used, the laser spectrum width is wider and can not resolve the energy gap between
	$\ket{1s2p~^3P}$ and $\ket{1s2p~^1P}$ states. In that case, two peaks merge at around 2.40 a.u., as shown in
	Fig. \ref{FS1}(b). 

	\begin{figure}
		\centering
		\includegraphics[width=0.48\textwidth]{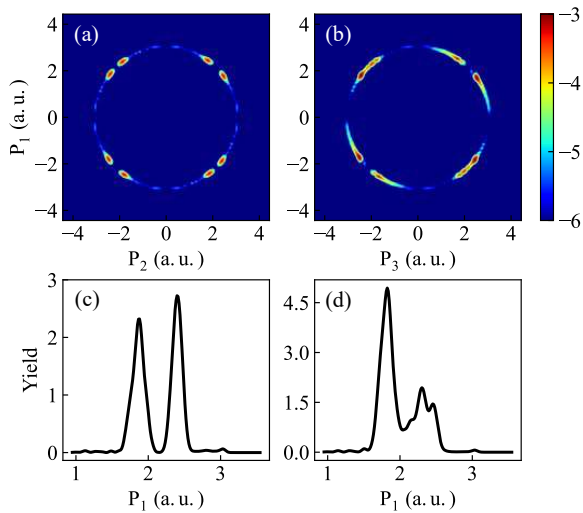}
		\caption{(a) (b) Correlated momentum spectra (in logarithmic scale) for two-photon DI (the ion stays in its first excited state)
			of $e_1^{\uparrow}e_2^{\uparrow}$ and $e_1^{\uparrow}e_3^{\downarrow}$, respectively.
			(c)(d) $e_1^{\uparrow}$ momentum distributions obtained by integrating the events in the second
			quadrant in (a) and (b) along the horizontal axis, respectively.
			The EUV pulse duration is 38 a.u..}
		\label{F2}
	\end{figure}

	\section{double ionization of Li}
	
	\subsection{Two-electron joint momentum distributions}	
	In the study of double ionization, we look into the joint momentum distribution
	(take $p_1$-$p_2$ as an example). We collect the wave function in the area $|x_1|>30$ a.u.,
	$|x_2|>30$ a.u. and $|x_3|<30$ a.u. at the end of the simulation $t=t_f$, and partially Fourier
	transform it with respect to $x_1$ and $x_2$ to obtain $\tilde{\Phi}_{1,2}(p_1,p_2,x_3;t_f)$.
	The projection $\langle p_1,p_2,\psi_n|\tilde{\Phi}_{1,2}(t_f)\rangle$ gives the amplitude
	that two electrons have the momentum $(p_1, p_2)$ and meanwhile Li$^{2+}$ is in the one-electron orbital $\psi_n(x_3)$.
	In this section, we resolve the sequential inner-shell double ionization paths for the electron pairs with different spin states and extract the channel induced by the orbital swap mechanism.

	Figures \ref{F2} (a) and (b) show the $e_1^{\uparrow}e_2^{\uparrow}$ and $e_1^{\uparrow}e_3^{\downarrow}$ joint momentum
	distributions for two-photon double ionization with Li$^{2+}$ in its first excited state, respectively.
	In both panels, the events that we collect are distributed on the circle
	$p_1^2/2+p_{2,3}^2/2=2\omega-I_p$, where $I_p$ is the energy threshold to free two inner-shell electrons.
	To show their differences more clearly, we integrate the events in Fig. \ref{F2}(a) and (b)
	along the horizontal axis to plot
	the momentum distribution of the electron $e_1^{\uparrow}$, as shown in Figs. \ref{F2}(c) and (d).
	Both curves present two clear peaks at around 1.85 a.u. and 2.4 a.u..
	The peaks at 2.4 a.u. in all panels are trivial, i.e., $e_1^{\uparrow}$ absorbs one photon and is ionized from the inner shell (see in Fig. \ref{FS1}).
	To produce the peak at 1.85 a.u. in Fig.\ref{F2}(d), $e_3^{\downarrow}$ is firstly ionized by
	absorbing a photon, and then the inner-shell $e_1^{\uparrow}$ is sequentially released by absorbing the
	second photon. This scenario can be derived from the energy diagrams of
	Li and Li$^+$ shown in Table \ref{table1}. The above analysis establishes
	that Li absorbs two photons sequentially.
	For DI of $e_1^{\uparrow}$ and $e_3^{\downarrow}$, the peak around 1.85 a.u. is much higher than
	that around 2.4 a.u. in Fig. \ref{F2}(d), which shows that $e_1^{\uparrow}$ has a larger probability
	to absorb the second photon, i.e., path-$\mathrm{\uppercase\expandafter{\romannumeral1}}$ 
	in Fig. \ref{F1} dominates. However, for DI of $e_1^{\uparrow}e_2^{\uparrow}$, the two electrons 
	have the same spin and are located on different shells initially,
	which seems to be contradictory to the previously mentioned mechanism requiring
	the photoionization of two inner-shell electrons. 
	A possible way to meet the energy diagram is that the inner-shell $e_3^{\downarrow}$ 
	absorbs the second photon and knocks out the outer-shell spin-up electron on its way out.
	However, such a pathway, similar to the well-known ``shake off" process, 
	has small cross section especially when the driving laser pulse is not too short. 
	This scenario can be excluded by numerical simulations, which are to be discussed  
	in Fig. \ref{FS3}.
	Hence, to simultaneously meet the spin and energy requirement, the outer-shell 
	electron must fill the hole in the inner shell in some manner
	once the spin-up inner-shell electron has been ionized by absorbing the first photon.
	Hence, the two electrons on two different shells in Li$^+$ must swap their orbitals which is 
	the so-called orbital swap mechanism
	in order to preserve the energy conservation after the single ionization.
	When the outer-shell spin-up electron hops to the inner shell, it can absorb another 
	photon, contributing to the peak
	around 1.85 a.u. in Fig. \ref{F2} (c). This scenario is described by
	path-$\mathrm{\uppercase\expandafter{\romannumeral3}}$ in Fig. \ref{F1},
	ending with the two photoelectrons in the spin-triplet state.
	
	\subsection{Streaking photoelectron spectra}
	
	\begin{figure}
		\centering
		\includegraphics[width=0.48\textwidth]{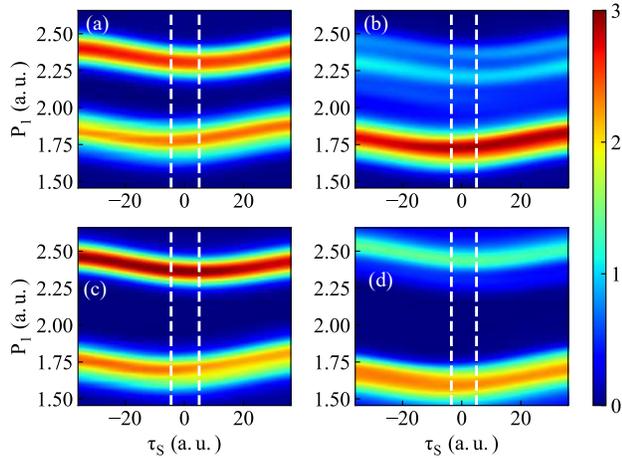}
		\caption{The streaking spectrogram of $e_1^{\uparrow}$ along the positive direction 
			under different conditions.  (a) and (b): Two electrons emit along the same direction.
			(c) and (d): Two electrons emit along opposite directions. The left and right columns
			are for the double ionization events of $e_1^{\uparrow}e_2^{\uparrow}$ 
			and $e_1^{\uparrow}e_3^{\downarrow}$, respectively. The one-cycle 1200 nm streaking 
			IR field has the intensity $5.0 \times 10^{11} \mathrm{~W} / \mathrm{cm}^{2}$.
			The white dashed curves mark the troughs of the streaked patterns.
		}
		\label{FS2}
	\end{figure}
	
	In the above section, we claim that Li absorbs two EUV photons sequentially. This statement 
	is confirmed by photoelectron energy spectra in Fig. \ref{F2}. However, the best way to support
	the sequential absorption of two photons is by investigating the streaking spectrogram of
	photoelectrons\cite{streaking,streaking-He}. In this subsection, we use the one-cycle (to ensure our simulation
	box can hold all the two-photon double ionization wavepackets) 1200 nm laser pulse with the intensity
	$5.0 \times 10^{11} \mathrm{~W} / \mathrm{cm}^{2}$ to streak the double ionization events.
	Approximately, the final photoelectron momentum under the streaking field can be written as
	\begin{equation}
		\vec{p}(t_f)=\vec{p}_0-\vec{A}_{s}(\tau_s+\Delta t_s),
		\label{streaking}
	\end{equation}
	where the streaking laser vector potential is
	\begin{equation}
		\vec{A}_{s}(\tau_s)=\vec{A}_0\sin^2[\pi(\tau_s+0.5T_{s})/T_{s}]\cos(\omega_{s}\tau_s).
		\label{streaking_a}
	\end{equation}
	Here, $\tau_s$ is the time delay between the center of the EUV pulse and the center of the
	streaking field, $T_{s}$ is the period of the streaking field.
	$\Delta t_s$ is the streaking time delay, and $\vec{p}_0$ is the photoelectron momentum when the
	streaking field is absent. By fitting the simulation results with Eq. (\ref{streaking}), we can get
	the streaking time delay. A negative (positive) time delay means that the photon is absorbed
	before (after) the center of the EUV pulse.
	In Figs. \ref{FS2}(a) and (c), we show the streaking spectrograms (DI of $e_1^{\uparrow}e_2^{\uparrow}$)
	for two electrons emitting along the same and opposite directions, respectively.
	Our simulation for both cases shows that the traces at
	the high momentum peak (around 2.4 a.u.) and the low momentum peak (around 1.85 a.u.)
	have time delays of about $-5.0$ a.u. and 4.5 a.u., respectively, as marked by the
	vertical dashed lines in each panel. These results confirm the sequential two-photon ionization scenarios.
	The time interval of absorbing the first and second photons relies on the EUV pulse duration \cite{streaking-He}.
	Different from the sequential double ionization
	of He in Ref \cite{streaking-He}, our two-photon absorption delay will also be
	influenced by the orbital swap mechanism. For instance, for DI of $e_1^{\uparrow}e_3^{\downarrow}$ ,
	the streaking delay of $e_1^{\uparrow}$ for the peak around 1.85 a.u. is about 3.5 a.u.,
	as shown in Fig. \ref{FS2} (b) and (d), which is smaller than the value in DI of $e_1^{\uparrow}e_2^{\uparrow}$.
	Such a difference can be explained as follows. For DI of $e_1^{\uparrow}e_2^{\uparrow}$,
	after $e_1^{\uparrow}$ is emitted, Li$^+$ needs some time to accomplish the orbit swap between $e_2^{\uparrow}$ and $e_3^{\downarrow}$.
	The second photon absorption occurs after $e_2^{\uparrow}$ and $e_3^{\downarrow}$ swap.
	The streaking time difference for DI of $e_1^{\uparrow}e_2^{\uparrow}$ and $e_1^{\uparrow}e_3^{\downarrow}$  has not been studied before
	and it carries the information of correlated electrons which deserves to be studied further.
	
	\subsection{Inelastic scattering in double ionization}
    According to the photoelectron momentum distribution and streaking spectrogram, it is
	the two inner-shell electrons that absorb two photons and get freed. It is intuitive to have two
	photoelectrons with opposite spins since the two inner-shell electrons automatically satisfy
	all requirements. However, obtaining two photoelectrons with the same spin sounds contradict
	to the initial condition since these two electrons distribute on two different shells. 
	Besides the orbital swap, another plausible way to meet the energy diagram is the following. 
	After $e_1^{\uparrow}$ is released, $e_3^{\downarrow}$ in the inner shell may absorb the 
	second photon. During its way out, $e_3^{\downarrow}$ knocks out the out-shell electron
	$e_2^{\uparrow}$ and itself is captured in the outer shell. In this way, the two photoelectrons
	are both spin up and their joint momentum distributions would be similar as shown in Fig. \ref{F2}. 
	However, in this section, we will show such a pathway only is nonnegligible when the
	driving laser pulse is extremely short. Usually, this pathway can be excluded due to very tiny probabilities.

	\begin{figure}
		\centering
		\includegraphics[width=0.48\textwidth]{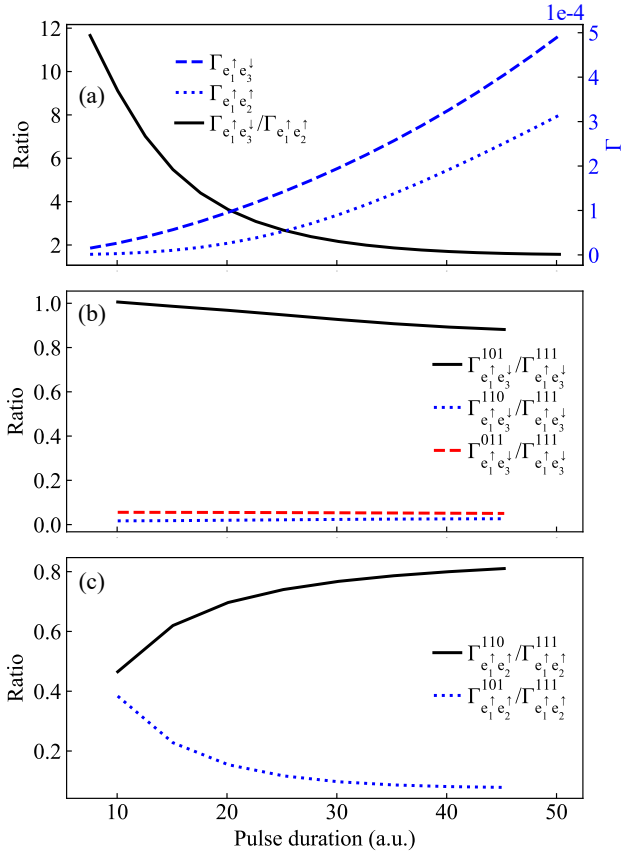}
		\caption{(a) The two-photon DI probability with Li$^{2+}$ in its first excited state as 
			a function of the lase pulse duration. The blue dashed and dotted lines are for 
			$e_1^{\uparrow}e_2^{\uparrow}$ and $e_1^{\uparrow}e_3^{\downarrow}$ , respectively. 
			The black solid line shows the ratio of them two. 
			(b) and (c): The double ionization probabilities obtained by selectively setting one of $f_i=0$
			as a function of the pulse duration.}
		\label{FS3}
	\end{figure}
	
	To explore how the photon energy deposits into Li, we artificially revise the laser-electron coupling term
	by adding three parameters ($f_1,f_2,f_3$) into the dipole interaction to turn on ($f_i=1$)
	or turn off ($f_i=0$) the interaction for a certain electron with the laser field	
	\begin{equation}
		W(t)=A(t)(f_1p_1+f_2p_2+f_3p_3).
	\end{equation}
	By investigating the $f_j$-dependent double ionization rate, one can estimate the importance of
	electron-electron Coulomb correlation in double ionization. 
	First, we include the full laser-electron coupling, i.e., $f_1=f_2=f_3=1$ , and show the ratio between the ionization rates of $e_1^{\uparrow}e_2^{\uparrow}$ and 
	$e_1^{\uparrow}e_3^{\downarrow}$ in Fig. \ref{FS3}(a). The sensitive dependence of the ratio on 
    the pulse duration indicates that the ionization mechanisms are different for different pulse 
    duration.
	Then, by turning off part of the laser-electron interaction, we can realize how important 
	of the laser-electron interaction, and on the contrary, the importance of the 
	complementary electron-electron correlation. Figures. \ref{FS3}(b) and \ref{FS3}(c)
	show the ionization probability as a function of the laser pulse duration.
	The three parameters marked in the superscript of the ionization probability
	($\Gamma^{f_1f_2f_3}$) are used to express the results under different laser-Li interactions.
	From Fig. \ref{FS3}(b), for DI of $e_1^{\uparrow}e_3^{\downarrow}$, the main double ionization channel is
	the photon absorption of the two inner-shell electrons  ($e_1^{\uparrow}$ and $e_3^{\downarrow}$),
	and other channels which need inelastic scattering between electrons
	have a negligible probability. This supports the main mechanism for DI of $e_1^{\uparrow}e_3^{\downarrow}$ (path-$\mathrm{\uppercase\expandafter{\romannumeral1}}$ in Fig. \ref{F1}).
	
	For DI of $e_1^{\uparrow}e_2^{\uparrow}$, as presented in Fig. \ref{FS3}(c),
	we find that the main channel originates from the photon absorption of 
	$e_1^{\uparrow}$ and $e_2^{\uparrow}$ when the pulse duration is not very short, 
	which also excludes the inelastic channel for these parameters.
	When the pulse duration is very short, the probability of the sequential photon absorption channel 
	for DI of $e_1^{\uparrow}e_2^{\uparrow}$ ($\Gamma^{110}_{e_1^{\uparrow}e_2^{\uparrow}}$) 
	has the same order of magnitude with the inelastic scattering channel 
	($\Gamma^{101}_{e_1^{\uparrow}e_2^{\uparrow}}$). However, the contribution of the inelastic 
	scattering channel is usually much smaller than the normal sequential two-photon process (path-$\mathrm{\uppercase\expandafter{\romannumeral1}}$), as shown by the black-solid curve
	in Fig. \ref{FS3}(a). With the increasing of the laser pulse duration, the proportion of the 
	inelastic scattering channel is smaller and smaller. 
	Based on our analysis above, we can deduce that the channels which need inelastic 
	scattering are not the main channel for two-photon DI with excitation in this work. 
	For the sequential photon absorption channel for DI of $e_1^{\uparrow}e_2^{\uparrow}$, 
	there should be a certain time delay between the sequential two-photon absorption to 
	wait for the orbit swap happens, and a very short laser pulse does not favor the mechanism.

\section{The orbital swap mechanism}
	
	\subsection{Formulating the orbital swap}
	\label{osw}
	Above simulation results all support the orbital swap of two electrons in Li$^+$ after single ionization. 
	In this subsection, we formulate the orbital swap. Physically, the ground state can be rewritten as \cite{cross-s2}
	\begin{equation}
		\begin{aligned}
			&\sqrt{2} \Phi_{\alpha \alpha \beta}\left(x_{1}, x_{2}, x_{3}\right)=\langle x_{1}, x_{3}|1 s^{2} \,^1S\rangle \langle x_{2}| 2 p\rangle\\
			&+\frac{1}{\sqrt{2}}\left[\langle x_{1}, x_{3}|1 s 2 p\,^{3} P_{M_{S}=0}\rangle
			-\langle x_{1}, x_{3}| 1 s 2 p\,^{1} P \rangle\right]\langle x_{2}| 1 s\rangle.
		\end{aligned}
		\label{eq_gs_e2}
	\end{equation}
	This expression can be confirmed by the single ionization spectra in Fig. \ref{FS1}.
	If the $1s$-electron $e_2^{\uparrow}$ is removed at $t_1$, the remaining Li$^+$ will be in the superimposed state consisting of the triplet state $\ket{A}$=$|$$1s2p$$\,$$^{3}P_{M_{S}=0}$$\rangle$ and the singlet state	$\ket{B}$=$|$$1s2p$$\,$$^{1}P$$\rangle$. One can expand the spatial wave functions of the two states with one-dimensional Li$^+$ orbitals $\phi_n$ as
	\begin{equation}
		\begin{aligned}
			\sqrt{2}\langle x_1,x_3\ket{A}=\phi_{1s}(x_1)\phi_{2p}(x_3)-\phi_{1s}(x_3)\phi_{2p}(x_1),\\
			\sqrt{2}\langle x_1,x_3\ket{B}=\phi_{1s}(x_1)\phi_{2p}(x_3)+\phi_{1s}(x_3)\phi_{2p}(x_1).
		\end{aligned}
		\label{eq_gs_AB}
	\end{equation}
	To complete the derivation, we assume the photoelectron is in plane wave
	orbitals $|\epsilon\rangle$ and rewrite the wave function of the (Li$^+$,e$^-$) system
	produced at $t_1$ after the emitting of $e_2^{\uparrow}$
	\begin{equation}
		\begin{aligned}
			|\Psi_{t_1}(t)\rangle=&\int d\varepsilon |\varepsilon\rangle e^{-{\mathrm{i}}\varepsilon (t-t_1)}\bigg\{|A\rangle C_{t_1}(E_A+\epsilon)e^{-{\mathrm{i}} E_A(t-t_1)}\\
			&+|B\rangle C_{t_1}(E_B+\epsilon)e^{-{\mathrm{i}} E_B(t-t_1)}\bigg\}.
		\end{aligned}
		\label{eq_DI}
	\end{equation}
	Here $E_A$ and $E_B$ are energies of $\ket{A}$ and $\ket{B}$
	states (see Table \ref{table1}), and $|C_{t_1}(E)|^2$ is the energy distribution centered at $E=E_{gs}+\omega$ ($E_{gs}$ is the ground state energy),
	which is determined mainly by the photon energy bandwidth.
	In the extreme case that the single ionization is triggered by a delta pulse at $t_1$,
	$C_{t_1}(E)$ would have infinite width and $C_{t_1}(E_A+\epsilon)\approx C_{t_1}(E_B+\epsilon)$.
	By projecting Eq. (\ref{eq_DI}) onto $|2p,\varepsilon,1s\rangle$ and integrating over $\varepsilon$, one can obtain the probability of the spin-down electron being in the inner-shell
	\begin{equation}
		P^{1s}_{\beta}(t)\propto \cos^2\left[\Delta E \ (t-t_1)/2\right]. \label{eq_tdob}
	\end{equation}
	Eq. (\ref{eq_tdob}) indicates that the period of orbital swap is $T=2\pi/\Delta E$.
	In our one-dimensional model, this period is about 24 a.u..
	We further emphasize that we neglect the laser-spin coupling in this study since it is much
	smaller. It is the mechanism of orbital swap that causes this oscillation and provides
	a unique opportunity to control the ionization pathways.
	Note that $\Delta E$ in our one-dimensional model is larger than that in the real
	three-dimensional atom, however, this model already explores the mechanism without a doubt.

	\subsection{Time-resolved orbital swap dynamics}
	
	\begin{figure}
		\centering
		\includegraphics[width=0.48\textwidth]{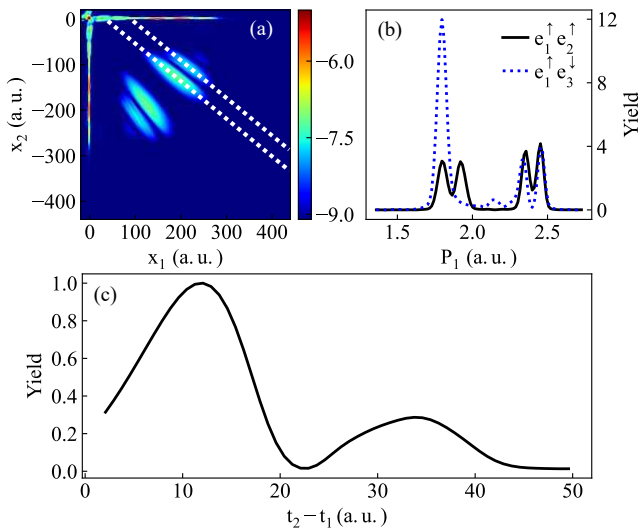}
		\caption{(a) The wave function distribution (in logarithmic scale) for DI (the ion stays in its first excited state)
			of $e_1^{\uparrow}e_2^{\uparrow}$ at $t=t_f$. (b) $e_1^{\uparrow}$ momentum spectra
			(normalized by the DI probability of $e_1^{\uparrow}e_2^{\uparrow}$) for DI
			of $e_1^{\uparrow}e_2^{\uparrow}$ (black solid) and $e_1^{\uparrow}e_3^{\downarrow}$
			(blue dotted). (c) The distribution of the time difference between absorbing the first
			and second photon. The pulse duration is 100 a.u..}
		\label{F3}
	\end{figure}
	
	If the driving laser pulse duration is longer than the orbital swap period, the two-electron swap
	may play a significant role in two-photon double ionization.
	To see that, we perform the simulation using a longer laser pulse ($\tau=100$ a.u.) and diagnose
	the double ionization for $e_1^{\uparrow}e_2^{\uparrow}$.
	Figure \ref{F3}(a) presents the wave function distribution at the end of the simulation for DI of 
	$e_1^{\uparrow}$ and $e_2^{\uparrow}$ associating with the Li$^{2+}$ in its first excited state.
	One can clearly see spatially separated wave packets guided
	by the white dashed curves. The intercepts of the white dotted
	line to $x_2=0$ are the propagation distances of the first ionized electron just
	before Li$^+$ absorbs the second photon. Since the momentum of the first ionized electron is given,
	the time interval between absorbing the first and second photon ($t_2-t_1$) can be
	numerically extracted easily, as shown in Fig. \ref{F3}(c). One may expect more peaks
	if the driving laser pulse duration is even longer. The maxima appear around
	$t_2-t_1=(j-\frac{1}{2})T$
	($j$ is a positive integer) when $e_3^{\downarrow}$ hops to the $n=2$ shell.
	This numerical result has a good quantitative agreement with the theoretical
	prediction. For DI of $e_1^{\uparrow}e_2^{\uparrow}$, the photon energy bandwidth
	is narrow enough to resolve $\Delta E$ and thus two subpeaks for $e_1^{\uparrow}$
	around 2.4 a.u. appear, as shown in Fig. \ref{F3} (b). These two subpeaks correspond to leaving Li$^+$ in
	the $|A\rangle$ and $|B\rangle$ states after absorbing the first photon.
	The two subpeaks at around 1.8 a.u. correspond to the ionization of the $1s$ electron of Li$^+$
	from the $|A\rangle$ and $|B\rangle$ superposition by absorbing the second photon.
	For DI of $e_1^{\uparrow}e_3^{\downarrow}$, the emission of $e_3^{\downarrow}$ only
	produces Li$^+$ in the spin-triplet state, and thus no subpeaks at 1.8 a.u. appear.
	
	\section{generating spin-polarized electron pairs}
	
	\begin{figure}
		\centering
		\includegraphics[width=0.48\textwidth]{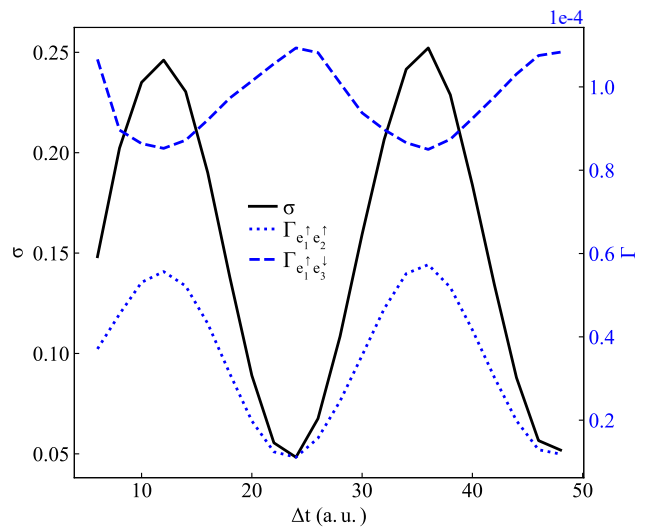}
		\caption{(a) The spin polarization $\sigma$ and the two-photon DI probabilities
			of $e_1^{\uparrow}e_2^{\uparrow}$ and $e_1^{\uparrow}e_3^{\downarrow}$ as a function of
			the time delay between two attosecond pulses. Each pulse duration is 10 a.u..
		}
		\label{F4}
	\end{figure}
	The orbital swap mechanism can help us control the ultrafast photoionization dynamics.
	Here, we show that the principle of the intrinsic orbital swap in Li$^+$ can be applied to produce
	spin-selective photoelectron pairs. Previous works showed that one can generate
	spin-polarized electron pairs through the electron impact \cite{ion-impact1,ion-impact2}.
	Here, we use two sequential attosecond pulses to generate the spin-polarized photoelectron pair in
	the double ionization of Li. The first attosecond pulse kicks off a spin-up
	inner-shell electron, leading to the two-electron orbital swap in Li$^+$, as described by Eq. (\ref{eq_tdob}).
	If the second attosecond pulse arrives when the spin-up electron hops to the inner
	shell, another spin-up photoelectron is emitted.
	Thus, we obtain the two correlated electrons in the triplet state. On the contrary, if the second
	attosecond pulse arrives when the spin-down electron swaps to the inner shell, one obtains
	two photoelectrons with opposite spins.
	In simulations, the laser field is composed of two identical attosecond pulses, and each one is expressed
	by Eq. (\ref{laser}). Each pulse duration is $\tau_{EUV}=10$ a.u., and the time delay between two
	attosecond pulses is variable.
	In Fig. \ref{F4}, we present the double ionization probabilities $\Gamma_{e_1^{\uparrow}e_2^{\uparrow}}$ and $\Gamma_{e_1^{\uparrow}e_3^{\downarrow}}$ with Li$^{2+}$ in its first excited state. We define the spin polarization as
	\begin{equation}		\sigma=\frac{\Gamma_{e_1^{\uparrow}e_2^{\uparrow}}}{\Gamma_{e_1^{\uparrow}e_2^{\uparrow}}+\Gamma_{e_1^{\uparrow}e_3^{\downarrow}}+\Gamma_{e_2^{\uparrow}e_3^{\downarrow}}}.
	\end{equation}
	Note that $\Gamma_{e_1^{\uparrow}e_3^{\downarrow}}=\Gamma_{e_2^{\uparrow}e_3^{\downarrow}}$ due to the identity of $e_1^{\uparrow}$ and $e_2^{\uparrow}$.
	The spin polarization is plotted by the black-solid curve
	in Fig. \ref{F4}. Both $\Gamma_{e_1^{\uparrow}e_2^{\uparrow}}$ and $\Gamma_{e_1^{\uparrow}e_3^{\downarrow}}$, as well as the spin polarization vary
	with the time delay, and the variational period is same as the two-electron swap period in Li$^+$.
	However, the total double ionization probability $\Gamma_{e_1^{\uparrow}e_2^{\uparrow}}+2\Gamma_{e_1^{\uparrow}e_3^{\downarrow}}$ keeps
	unchanged.
	
	\begin{figure}
		\centering
		\includegraphics[width=0.48\textwidth]{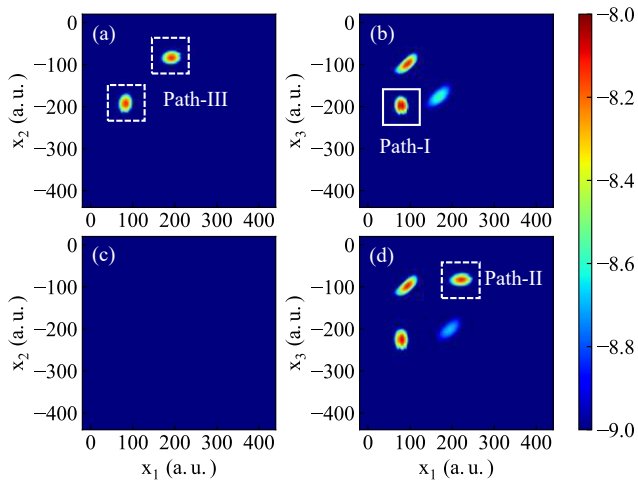}
		\caption{The wave function distributions (in logarithmic scale) at the end of the calculations
			when the time delays between two attosecond pulses are (a, b) $\Delta t=36$ a.u. and (c, d)
			$\Delta t=48$ a.u.. The left and right columns are for DI of $e_1^{\uparrow}e_2^{\uparrow}$ and $e_1^{\uparrow}e_3^{\downarrow}$, respectively.
			The duration of each pulses is 10 a.u., the central frequency is $\omega=5$ a.u., 
			and the intensity is $4.0 \times 10^{16} \mathrm{~W} / \mathrm{cm}^{2}$.}
		\label{FS4}
	\end{figure}
	
	To better understand the double ionization dynamics, in Fig. \ref{FS4}, we show the snapshots of the final double ionization with excitation wave function distributions when
	the time delays between the two attosecond pulses are 36 a.u. (upper row) and 48 a.u.(lower row).
	The left and right columns present DI of $e_1^{\uparrow}e_2^{\uparrow}$ 
	and $e_1^{\uparrow}e_3^{\downarrow}$, respectively. In Fig. \ref{FS4}(a), the wave packets come from path-$\mathrm{\uppercase\expandafter{\romannumeral3}}$. In Fig. \ref{FS4}(b), the wave packets along the diagonal come from the direct
	emitting of the two inner-shell electrons by absorbing two photons from the single pulse.
	The wave packet in the left lower corner comes from path-I. The main signals at this
	time delay (36 a.u., half integrals of the period) will not be affected by the time delay and we can regard them
	as the background signals for DI of $e_1^{\uparrow}e_3^{\downarrow}$ which limit the maximum value of
	the spin polarization parameter $\sigma$. When the delay is 48 a.u., only path-$\mathrm{\uppercase\expandafter{\romannumeral1}}$ and path-$\mathrm{\uppercase\expandafter{\romannumeral2}}$ exist, thus,
	the double ionization probability for $e_1^{\uparrow}e_2^{\uparrow}$ is too small to be visible, as shown
	in Fig. \ref{FS4}(c). Path-$\mathrm{\uppercase\expandafter{\romannumeral1}}$ and path-$\mathrm{\uppercase\expandafter{\romannumeral2}}$ contribute to the signals in Fig. \ref{FS4}(d). By changing the time delay, the ratio of the ionization probabilities between path-$\mathrm{\uppercase\expandafter{\romannumeral2}}$ ($e_1^{\uparrow}e_3^{\downarrow}$) path-$\mathrm{\uppercase\expandafter{\romannumeral3}}$ ($e_1^{\uparrow}e_2^{\uparrow}$) can be controlled.
	
	\section{conclusion and outlook}
	\label{xxx}
	In summary, we reveal an unexplored orbital swap mechanism by investigating the double ionization of Li in attosecond EUV pulses.
	Depending on the photoelectron spin orientation in single ionization of Li, 
	Li$^+$ may be in the superposition with spatial exchange symmetry and antisymmetry. Such a quantum beat forces
	the two bound electrons in Li$^+$ to swap their orbitals periodically. 
	By characterizing and precisely timing the ultrafast orbital swap dynamics,
	one may directly control the spin polarization of the photoelectron pair by the attosecond-pump
	attosecond-probe strategy \cite{Chini14,Kang20}. 
	
	The difficulty of directly observing the orbital swap process lies on measuring both photoelectron spin orientations
	simultaneously on top of the COLTRIMS measurement, which is a technology still under developing. However, this
	difficulty can be partly relieved by only measuring one photoelectron spin.
	In the two-photon sequential double ionization, the secondly released electron has lower energy compared to the 
	firstly released electron. As discussed in Section \ref{osw}, not only the spin polarization of the electron
	pair, but also the spin polarization of the secondly released electron oscillates with the time delay.
	Therefore, measuring the spin polarization of the photoelectron with lower energies by the TOF-Mott spectrometer \cite{circular4} will give a proof of orbital swap in Li$^+$.

	The orbital swap mechanism can even be confirmed without measuring any photoelectron spin. Since the orbital
	swap will inevitably determine other process, for example, the high harmonic generation, we can retrieve the 
	orbtial swap from the high harmonic spectra. For example, exposing Li in combined EUV and midinfrared laser pulses,
	an inner-shell spin-up electron is knocked out by the EUV pulse and then driven back to Li$^+$. 
	The rescattering process may be severely modulated by the orbital swap mechanism. 
	If the photoelectron excursion time is half integers of the swap period, the dipole is annihilated and no harmonics
	emit. By diagnosing the fine structures in the harmonic spectra, one is possible to retrieve the ultrafast orbital swap. 
	The orbital swap mechanism discovered in this study advances our understanding of atoms with open-shell structures,
	and more ultrafast spin-resolved dynamics in multi-electron systems can be explored in the future.

\section*{Acknowledgements}
	This work was supported by National Natural Science Foundation of China (NSFC) (Grants No. 11925405),
	National Key R\&D Program of China (2018YFA0404802), and National Natural Science Foundation of
	China (NSFC) (Grants No. 91850203).
	Simulations were performed on the $\pi$ supercomputer at Shanghai Jiao Tong University.
	
	\bibliographystyle{unsrt}

\end{document}